\def\add#1{{\textcolor{black}{#1}}}

\documentclass[aps,prl,twocolumn,superscriptaddress,showpacs]{revtex4-1}
\usepackage[colorlinks,linkcolor=blue,citecolor=blue,urlcolor=blue,breaklinks=true]{hyperref}
\usepackage{amsmath,amssymb,graphics,epsfig,epstopdf,color,verbatim,braket,tabularx,breakurl}
\usepackage{multirow}
\usepackage{diagbox}
\usepackage{subfig}
\usepackage[normalem]{ulem}
\usepackage{float}
\usepackage{ragged2e}
\usepackage{graphicx}

\begin{document}

\title{A hybrid method integrating Green's function Monte Carlo and projected entangled pair states}

\author{He-Yu Lin}
\address{Department of Physics, Renmin University of China, Beijing 100872, China}
\address{Key Laboratory of Quantum State Construction and Manipulation (Ministry of Education), Renmin University of China, Beijing, 100872, China}
\author{Rong-Qiang He}
\address{Department of Physics, Renmin University of China, Beijing 100872, China}
\address{Key Laboratory of Quantum State Construction and Manipulation (Ministry of Education), Renmin University of China, Beijing, 100872, China}
\author{Yibin Guo}
\address{Deutsches Elektronen-Synchrotron DESY, Platanenallee 6, 15738 Zeuthen, Germany}
\author{Zhong-Yi Lu}
\email{zlu@ruc.edu.cn}
\address{Department of Physics, Renmin University of China, Beijing 100872, China}
\address{Key Laboratory of Quantum State Construction and Manipulation (Ministry of Education), Renmin University of China, Beijing, 100872, China}
\address{Hefei National Laboratory, Hefei 230088, China}
\date{\today}

\begin{abstract}
This paper introduces a hybrid approach combining Green's function Monte Carlo (GFMC) method with projected entangled pair state (PEPS) ansatz.
This hybrid method regards PEPS as a trial state and a guiding wave function in GFMC. By leveraging PEPS's proficiency in capturing quantum state entanglement and GFMC's efficient parallel architecture, the hybrid method is well-suited for the accurate and efficient treatment of frustrated quantum spin systems.
As a benchmark, we applied this approach to study the frustrated $J_1$-$J_2$ Heisenberg model on a square lattice with periodic boundary conditions (PBC).
Compared with other numerical methods, our approach integrating PEPS and GFMC shows competitive accuracy in the performance of ground-state energy. This paper provides systematic and comprehensive discussion of the approach of our previous work~\cite{Lin2024}.
\end{abstract}

\maketitle

\section{Introduction}
As a prominent research area in many-body physics, the intricate phenomena in two-dimensional quantum systems have garnered substantial interest. In many cases, the models describing these systems have strong interaction among particles, which makes analytical methods difficult and leads to the exponential wall problem in the exact diagonalization (ED). These problems stimulate the development of several numerical approximation methods. Among the various methods, quantum Monte Carlo (QMC) and tensor network states (TNS) have drawn increasing attention due to their success over the past decades~\cite{trivediGFMC,verstraete2008matrix,orus2019tensor,cirac2021matrix,banuls2023tensor,xiang2023density,nightingale1998quantum,sorella2017book}.

QMC excels in handling the thermal properties and phase transitions of higher-dimensional systems and it utilizes random sampling to handle the state simulation of quantum many-body systems. The Green's function Monte Carlo (GFMC)~\cite{trivediGFMC}, a member of the QMC family, starts from a prescribed trial wave function and utilizes imaginary-time evolution to estimate the ground-state energy. The fixed-node approximation~\cite{Haaf1995PRB} is employed during the evolution to circumvent the sign problem~\cite{Sorella2000PRB,Reynolds1982JCP,sorella2017book}. Physical observables, including correlation functions, can be efficiently obtained via importance sampling, with accuracy guaranteed if the trial wave function provides sufficiently accurate nodal information of the ground state.

TNS provides a faithful and sign-problem-free approach to representing quantum states that obey the entanglement area law, a widespread property of the ground state and the low-energy excited states of local gaped Hamiltonians~\cite{verstraete2008matrix,Mont2018,orus2019tensor,cirac2021matrix,banuls2023tensor,xiang2023density}. The projected entangled pair state (PEPS)~\cite{verstraete2004renormalization}, a widely used TNS ansatz in two-dimensional quantum systems, can capture the entanglement properties intrinsic to quantum states. However, contracting the tensor network with a general observable usually faces a leading computational cost of $\mathcal{O}(D^{12})$~\cite{orus2014practical, NTN2017, iTEBD2023} with bond dimension $D$. This high computational cost in contraction is challenging to address with multi-core CPUs and GPUs due to its iterative structures~\cite{orus2014practical,Roman2014EPJB,cirac2021matrix,banuls2023tensor,xiang2023density} and remains one of the main obstacles for the applications in many problems.

By integrating the advantages of GFMC~\cite{trivediGFMC} and PEPS~\cite{MSL-PRB2000, Clark-arXiv2014, Sebastian-PRB2014, MPQin-PRB2020, CPB2022}, we propose a hybrid method tailored for two-dimensional quantum systems. This method uses PEPS to produce accurate nodal information of the ground state, serving as a trial wave function for GFMC. As a byproduct, the contraction of PEPS from different physical configurations can be parallelized by importance sampling the contribution. This parallelization speeds up the contraction, reducing the leading computational cost to $\mathcal{O}(D^6)$ for each configuration when contracting a tensor network with bond dimension $D$.

To benchmark the validity and efficiency of this hybrid approach, we study the frustrated $J_1$-$J_2$ Heisenberg model on a square lattice with periodic boundary conditions (PBCs). The simulation focuses on the intermediate regime $J_2/J_1 \sim 0.5$, where the phase is still under debate~\cite{zhang2003valence,mezzacapo2012ground,jiang2012spin,Hu2013PRB,morita2015quantum,wang2018critical,ferrari2020gapless,liu2022gapless,Wang2013PRL,Liu2018PRB}. Compared with other numerical methods\cite{Schulz1994Magnetic,Choo2019PRB,Nomura2021PRX}, the hybrid approach shows competitive accuracy in determining the ground-state energy. The excellent energy performance indicates the effectiveness of this hybrid method. The rest of the paper is organized as follows. We introduce the GFMC method and PEPS in Section 2 and Section 3, respectively. Section 4 details the hybrid method combining PEPS and GFMC. The benchmark results for the $J_1$-$J_2$ Heisenberg model are compared with those from other numerical methods in Section 5. This paper closes with a summary in Section 6.

\section{Green's function Monte Carlo method\label{sec:gfmc}}

The Green's function Monte Carlo (GFMC) method uses the power method to filter out the ground state of quantum many-body systems~\cite{Anderson1975JCP,sorella2017book}. For any initial state $\left|{\Psi _{\rm{init}}}\right\rangle$ having non-zero overlap with the ground state $\left|{\Upsilon _0}\right\rangle$, after a sufficiently long imaginary time $\beta$ it will give the ground state,

\begin{equation}
\label{eq:ITP}
\mathop {\lim }\limits_{\beta \to \infty } {e^{ - \beta H}}\left| {\Psi _{\rm{init}}} \right\rangle  \propto \left| {{\Upsilon_0}} \right\rangle.
\end{equation}

In practical applications, the imaginary time $\beta$ is sliced into $M$ equal parts with fixed interval $\Delta \beta = \frac{\beta}{M}$. After inserting the completeness relation $\sum\limits_{x} {\left|x\rangle\langle x\right|} = 1$ in each time slice, the realization of Eq.~\eqref{eq:ITP} is identical to repeating $\Psi \left( {x',{\beta _0} + \Delta \beta } \right) = \sum\limits_{x} {G_{x',x}} \Psi \left( {x,{\beta _0}} \right)$ for $M$ times, where Green's function ${G_{x',x}}=\left\langle x' \right|{e^{ - \Delta \beta H}}\left| {x} \right\rangle$ is introduced.
Note that $\sum\limits_{x'} {G_{x',x}} \ne 1$ in general. We therefore divide ${G_{x',x}}$ into the product of two factors, that is ${G_{x',x}} = {p_{x',x}}{b_x}$ by introducing the normalization factor ${b_x} = \sum\limits_{x'} {G_{x',x}}$ and the transition probability ${p_{x',x}=\frac{{G_{x',x}}}{b_x}}$.
We use a Markov process to perform the imaginary-time evolution, in which a walker $\left({x},{w}\right)$ being dependent on the configuration $x$ and the weight $w$ of $x$ is propagated.
The ground-state energy is
\begin{equation}
\label{eq:E0}
    E_0 \approx \frac{{\left\langle {\left\langle {{w_n}{e_{{\rm{loc}}}}\left( {{x_n}} \right)} \right\rangle } \right\rangle }}{{\left\langle {\left\langle {{w_n}} \right\rangle} \right\rangle }},
\end{equation}
where $\left\langle \left\langle...\right\rangle\right\rangle$ represents the statistical averaging of all samples, $e_{\rm{loc}}\left( {{x_n}} \right) = \sum\limits_{x'} {{H_{x',{x_n}}}}$ is the local energy, and $x_n$ is the $n$th sample.
\add{For clarity, it should be noted that the accuracy of the ground-state energy depends solely on the length of the imaginary time, whereas the statistical error in the energy is determined by the number of samples.
In Monte Carlo processes, the statistical error $\sigma$ typically decreases as the square root of the number of samples $N$, following $\sigma \propto 1/\sqrt{N}$.}

Importance sampling is utilized to enhance the efficiency of the Markov process, which needs to include a suitable guiding wave function. In this study, we choose a PEPS, $\left|\Psi _{\rm{P}}\right\rangle$, as the guiding wave function. The more precise nodal information is captured by the guiding wave function, the closer result of the evolution is to the exact ground state.
For the state $\left|\Psi_{\rm{P}}\right\rangle$, a configuration $x$ satisfying the condition $\Psi_{\rm{P}}\left(x\right) \equiv \langle x| \Psi_{\mathrm{P}}\rangle = 0$ is a node of $\left|\Psi _{\rm{P}}\right\rangle$. All these nodes constitute the nodal surface and give the nodal information of $\left|\Psi _{\rm{P}}\right\rangle$. Detailed theoretical and technical aspects are discussed in Section 5.

As ${G_{x',x}}$ is a transition matrix, the off-diagonal matrix elements of ${G_{x',x}}$ must be non-negative, otherwise a sign problem arises. In the case of lattice models, the fixed-node approximation~\cite{Sorella2000PRB, Reynolds1982JCP, Haaf1995PRB} is implemented to circumvent the sign problem by introducing an effective Hamiltonian that is sign-problem free,

\begin{equation}
\label{eq:Heff}
\begin{split}
H_{x,x'}^{\rm{eff}} = \left\{ {\begin{array}{*{20}{c}}
{{H_{x,x}} + \left( {1 + \chi } \right){V_{\rm{sf}}}\left( x \right)~~~~{\rm{for}}~x' = x,}\\
{{H_{x,x'}}~~~~~~~~~~~~{\rm{for}}~x' \ne x,~{s_{x,x'}} < 0,}\\
{ - \chi {H_{x,x'}}~~~~~~~~{\rm{for}}~x' \ne x,~{s_{x,x'}} > 0.}
\end{array}} \right.
\end{split}
\end{equation}
where $\chi \ge 0$, and we specifically set $\chi = 0$ in this paper. The off-diagonal elements in the effective Hamiltonian are adjusted by these signs of ${s_{x,x'}} = \Psi _{\rm{P}}\left( x \right){H_{x,x'}}\Psi _{\rm{P}}\left( {x'} \right)$, ${V_{\rm{sf}}}\left( x \right) = \sum\limits_{x':{s_{x,x'}} > 0} {{H_{x,x'}}\frac{{\Psi _{\rm{P}}\left( {x'} \right)}}{{\Psi _{\rm{P}}\left( x \right)}}}$ is the local potential introduced to reduce energy, and ${x':{s_{x,x'}} > 0}$ means for $x'$ that satisfy ${s_{x,x'}} > 0$.
The effective ground-state energy is
\begin{equation}
    E_0^{{\rm{eff}}} = \frac{{\left\langle {\Upsilon _0^{{\rm{eff}}}} \right|{H^{{\rm{eff}}}}\left| {\Upsilon _0^{{\rm{eff}}}} \right\rangle }}{{\left\langle {{\Upsilon _0^{{\rm{eff}}}}}
 \mathrel{\left | {\vphantom {{\Upsilon _0^{{\rm{eff}}}} {\Upsilon _0^{{\rm{eff}}}}}}
 \right. \kern-\nulldelimiterspace}
 {{\Upsilon _0^{{\rm{eff}}}}} \right\rangle }},
\end{equation}
where $\left| {\Upsilon _0^{{\rm{eff}}}} \right\rangle$ is the equilibrium state obtained through the GFMC method for $H^{\rm{eff}}$, which can be considered as an approximation of the exact ground-state energy $E_{\rm{exact}}$ of the original Hamiltonian $H$.

\section{Projected entangled pair state\label{sec:peps}}

\begin{figure*}
	\centering
	\includegraphics[scale=0.22]{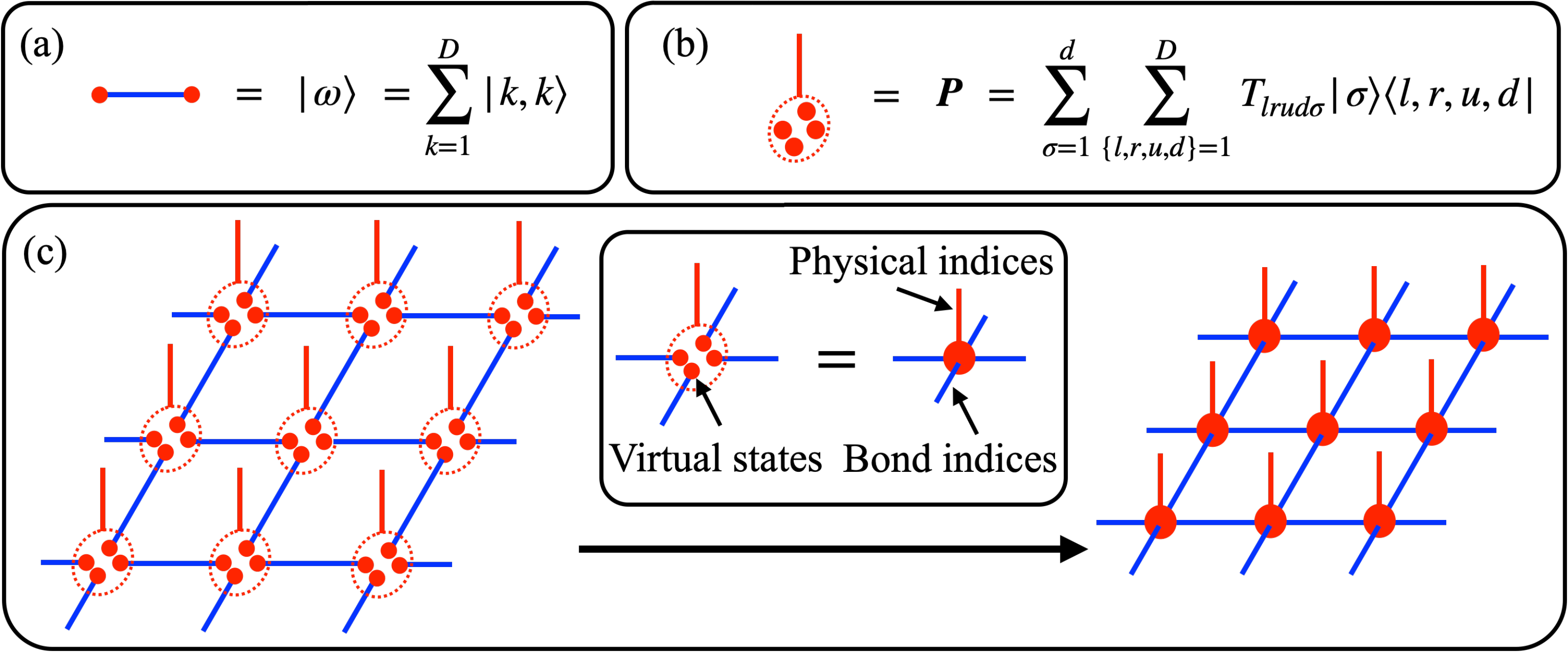}
	\caption{The formulation of a PEPS on a square lattice. A PEPS is constructed by (a) collecting all maximally entangled states $|\omega\rangle$ defined on the neighboring virtual sites, and (b) utilizing the projector $\boldsymbol{P}$ to map the four independent virtual states $\{|l\rangle,|r\rangle,|u\rangle,|d\rangle\}$ inside the same lattice site to physical states $|\sigma\rangle$. With the diagrammatic notation, the red dots represent the virtual states while the blue (vertical red) lines mark the virtual bond (physical) indices corresponding to their degree of freedom. (c) A two-dimensional many-body state is obtained by applying all projectors on the corresponding virtual states. Local tensors $T$ can be used as variational parameters in calculation. More details can be found in references~\cite{verstraete2004renormalization, Vidal2004PRL,verstraete2008matrix}.}
	\label{fig:PEPS}
\end{figure*}

This section starts by constructing a general PEPS on a square lattice. A PEPS is generally formed by combining maximally entangled states at every neighbouring virtual site and linear maps from virtual states to the physical states of each lattice site~\cite{verstraete2004renormalization, Vidal2004PRL}. As shown in Fig.~\ref{fig:PEPS}(c), each lattice site has four auxiliary virtual sites connected respectively to their nearest neighboring virtual sites. The connection represents the maximally entangled state $|\omega\rangle=\sum_{k=1}^D |k, k\rangle$, with $|k\rangle$ being the states $\{|1\rangle , |2\rangle, \cdots, |D\rangle\}$ on virtual sites, shown in Fig.~\ref{fig:PEPS}(a). Since the virtual sites in the same lattice site are independent, we need a general linear projector
\begin{equation}
    \boldsymbol{P} = \sum_{\sigma=1}^{d} \sum_{\{l,r,u,d\}=1}^{D} T_{l r u d \sigma} | \sigma \rangle \langle l, r, u, d |
\end{equation}
that maps virtual states $\{|l\rangle,|r\rangle,|u\rangle,|d\rangle\}$ (four red dots in Fig.~\ref{fig:PEPS}(b)) with dimension $D$ to physical states $|\sigma\rangle$ with dimension $d$. This formulation results in a rank-$5$ tensor $T$ with dimension $D\times D\times D \times D \times d$. Therefore, a PEPS ansatz $|\psi\rangle$, represented with the diagrammatic notation in the rightmost diagram of Fig.~\ref{fig:PEPS}(c), read as
\begin{equation}
	|\psi\rangle = \sum_{\{\sigma\}}\left(\mathrm{Tr}\prod_{i}T_{l_i r_i u_i d_i \sigma_i}^{(i)}\right)\big|\sigma_1\sigma_2\cdots\sigma_i\cdots\big\rangle.
	\label{Eq:PEPS}
\end{equation}
The subscript (superscript) $i$ is added to the indices of the state $(l_i,r_i,u_i,d_i)$ (the tensor $T$) to accommodate their site-dependent properties and run over all lattice sites. The trace ''$\mathrm{Tr}$'' in Eq.~\eqref{Eq:PEPS} sums over from $1$ to $D$ of all the bond indices. For a given spin configuration $\{\sigma_i\}$, this trace directly gives the coefficient. For lattice with $N$ sites, the total number of variational parameters of PEPS is $NdD^4$ if only real tensors are considered. This fact indicates that $D$, the bond dimension, is the key parameter to tune the performance of the PEPS ansatz.

From the above construction process, a direct conclusion is that PEPS naturally obeys the area law for the entanglement entropy. Any bipartition of the physical degrees of freedom will break the maximally entangled states, with the number of breaks proportional to the length of the separating boundary. Additionally, PEPS will capture more entanglement with increasing bond dimension $D$, as the more the maximally entangled states encode, the more entanglement is in that case.

\section{The technical details of the hybrid approach\label{sec:hybrid}}

\begin{figure*}
	\centering
	\includegraphics[scale=0.22]{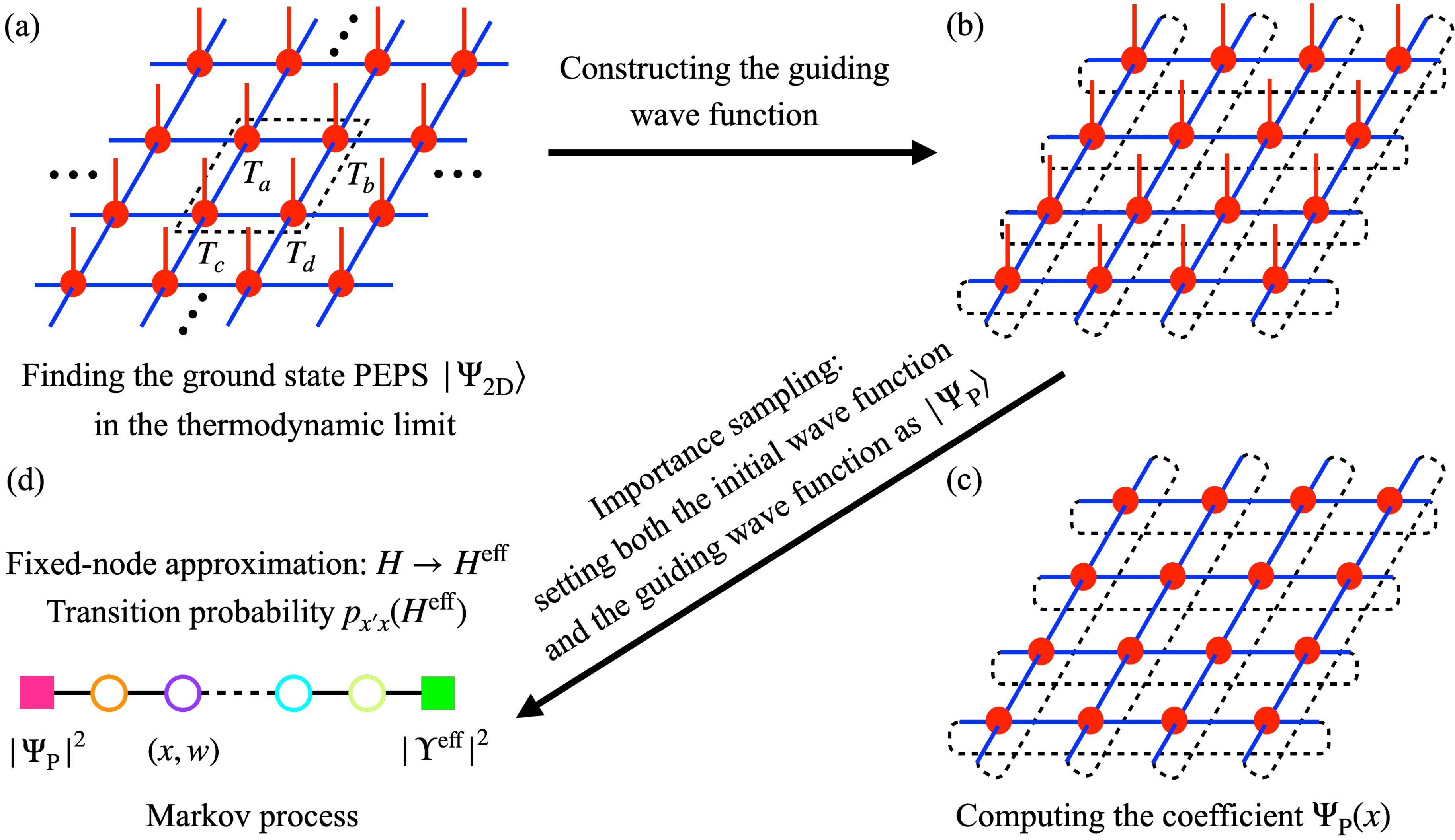}
	\caption{The flowchart of the hybrid approach. (a-b) Constructing the guiding wave function $|\Psi_{\mathrm{P}}\rangle$ with PEPS on finite size lattice and the imaginary-time evolution with GFMC from $|\Psi_{\mathrm{P}}\rangle$. $|\Psi_{\mathrm{P}}\rangle$ is obtained by replicating the tensors from the ground state PEPS $|\Psi_{\mathrm{2D}}\rangle$ on a square lattice in the thermodynamic limit. (c) The computation of the coefficient $\Psi_{\mathrm{P}}(x)$ for a given configuration $|x\rangle$, which is identical to contracting a two-dimensional tensor network with bond dimension $D$. (d) In GFMC, a walker evolves on the Markov chain from the initial distribution $|\Psi_{\rm{P}}|^2$ to the final equilibrium distribution $|\Upsilon^{\rm{eff}}|^2$. The walker is evolved by the transition probability $p_{x',x}$, which is a function of the effective Hamiltonian $H^{\rm{eff}}$ introduced in the fixed-node approximation.}
	\label{fig:Flow}
\end{figure*}

In this section, we present the detailed process of the hybrid method. This method includes two main parts, i.e., constructing a trial state from PEPS and continuing to perform the imaginary-time evolution for the state with GFMC. For convenience, we present the method by providing the details of models defined on a square lattice with periodic boundary conditions (PBCs). For models involving other lattices such as the Kagome lattice and triangular lattice, a similar procedure can be devised.

\subsection{Constructing a guiding wave function with PEPS\label{subsec:guidingPEPS}}
Let us start by constructing the guiding wave function from PEPS. Without loss of generality, we illustrate with a $4\times4$ square lattice with PBCs. Instead of directly optimizing the PEPS $|\Psi_{\mathrm{P}}\rangle$ on the finite lattice, we take a detour by considering the local tensors of the PEPS ground state $|\Psi_{\mathrm{2D}}\rangle$ in the thermodynamic limit as the approximation of local tensors on the finite lattice. This state preparation choice leads us to a three-step procedure. Firstly, shown in Fig.~\ref{fig:Flow}(a), is to find the PEPS with supercells by the well-developed algorithms of PEPS~\cite{xiang2023density}. \add{We choose a supercell with size $2\times2$, which is framed by a dashed parallelogram in Fig.~\ref{fig:Flow}(a) and is the smallest size to distinguish the four possible phases of the model. There are four distinct tensors in the supercell, say $\{T_a, T_b, T_c, T_d\}$, each of which is a tensor with shape $D\times D\times D\times D\times 2$ and initialized randomly.}
Then we adopt the differentiable programming tensor networks~\cite{WL2019PRX} to generate the ground-state PEPS $|\Psi_{\mathrm{2D}}\rangle$ with a $2\times2$ supercell in the thermodynamic limit. \add{$|\Psi_{\mathrm{2D}}\rangle$ is constructed by duplicating the supercell periodically. From the arbitrarily initialized $\{T_a, T_b, T_c, T_d\}$, one can calculate the energy $E=\langle \Psi_{\mathrm{2D}}|H|\Psi_{\mathrm{2D}}\rangle/\langle\Psi_{\mathrm{2D}}|\Psi_{\mathrm{2D}}\rangle$, then an automatic differentiation (AD) package can effectively obtain their gradients, i.e. $\{\frac{\partial E}{\partial T_a},\frac{\partial E}{\partial T_b},\frac{\partial E}{\partial T_c},\frac{\partial E}{\partial T_d}\}$, which can be used, e.g. in the stochastic gradient descent or the L-BFGS method, to update the local tensors $\{T_a, T_b, T_c, T_d\}$, and $|\Psi_{\mathrm{2D}}\rangle$ is thus updated in the direction of lower energy. This optimization procedure can be repeated until convergence is reached, and then we obtain an approximate PEPS representation $|\Psi_{\mathrm{2D}}\rangle$ of the ground state wave function.} Finally we replicate and periodically arrange the local tensors within the supercell to construct the trial function on the finite lattice, as shown in Fig.~\ref{fig:Flow}(b). Considering the balance between performance and computational cost, the bond dimension is constrained to be not greater than $7$.

After obtaining the guiding wave function $|\Psi_{\mathrm{P}}\rangle$, we continue to do importance sampling in the GFMC process. This requires us to compute the coefficient $\Psi_{\mathrm{P}}(x) \equiv \langle x| \Psi_{\mathrm{P}}\rangle$ for a given configuration $|x\rangle \equiv |x_1 x_2 \cdots x_i \cdots \rangle$. The computation of $\Psi_{\mathrm{P}}(x)$ is exactly the contraction of the two-dimensional tensor network with bond dimension $D$, i.e. $\mathrm{Tr}\prod_{i}T_{l_i r_i u_i d_i x_i}^{(i)}$, as shown in Fig.~\ref{fig:Flow}(c). Many tensor network contraction algorithms can be used to estimate $\Psi_{\mathrm{P}}(x)$ to a high precision~\cite{xiang2023density}. In our implementation, the higher-order tensor renormalization group (HOTRG)~\cite{Xie2012HOTRG} along one direction is adopted.

\subsection{Performing the imaginary-time evolution with GFMC\label{subsec:GFMC}}

This subsection introduces the procedures of performing the imaginary-time evolution from $|\Psi_{\mathrm{P}}\rangle$.
Two essential techniques are involved here: importance sampling and the fixed-node approximation.
The GFMC involves two wave functions: the initial trial wave function $\left|\Psi_{\rm{init}}\right\rangle$ for the evolution shown as Eq.~\eqref{eq:ITP}, and a guiding wave function for importance sampling.
In practice, we take $\left|\Psi_{\rm{P}}\right\rangle$ as the wave functions in both cases and thus the distribution of $x$ on the Markov chain follows $\left| \Psi_{\rm{P}}(x)\right|^2$ at the start of the evolution, as shown in Fig.~\ref{fig:Flow}(d), which is more efficient than starting from an arbitrary distribution. The expression of the ground-state energy in Eq.~\eqref{eq:E0} is changed to,

\begin{equation}
\label{eq:E0IM}
{E_0} \approx \frac{{\left\langle {\left\langle {{w_n}{{\tilde e}_{{\rm{loc}}}}\left( {{x_n}} \right)} \right\rangle } \right\rangle }}{{\left\langle {\left\langle {{w_n}} \right\rangle } \right\rangle }},
\end{equation}
where ${{\tilde e}_{\rm{loc}}}\left( {{x_n}} \right) = \sum\limits_x {{H_{x,x_n}}\frac{{\Psi_{\rm{P}} \left( {x} \right)}}{{\Psi_{\rm{P}} \left( x_n \right)}}}$ is the local energy taking into account importance sampling.

The fixed-node method is employed to overcome the sign problem. As described in Section 2, the original Hamiltonian $H$ which has a sign problem is replaced by ${H^{\rm{eff}}}$, which is sign-problem free and suitable for the GFMC method. It can be demonstrated that the energy ${E_0^{\rm{eff}}}$ obtained through the GFMC process for ${H^{\rm{eff}}}$ lies between the exact ground-state energy $E_{\rm{exact}}$ and ${E_{\rm{P}}}$ which is the observable of the original $H$ under the distribution $\left| \Psi_{\rm{P}} \right|^2$, i.e., ${E_0^{\rm{eff}}} \in [E_{\rm{exact}}, E_{\rm{P}}]$.
During the evolution process, the wave function $\left|\Psi^{\rm{eff}} \right\rangle$ described by the samples at each Markov-chain step maintains the same nodal surface as the guiding wave function $\left|\Psi_{\rm{P}}\right\rangle$, i.e., $\Psi_{\rm{P}}\left(x\right)$ and $\Psi^{\rm{eff}}\left(x\right)$ always have the same sign for any $x$. The evolution process is visually explained in Fig.~\ref{fig:Flow}(d). The horizontal lines represent the Markov chain, and the walker $\left(x, w\right)$ moves along the chain. The solid squares at both ends indicate that the system is in a stable distribution. The hollow circles in the middle, with different colors, represent the changing distribution of the system for each step. Since the Hamiltonian has been replaced, the transition probabilities are functions of ${H^{\rm{eff}}}$.

We give some comments about the computational cost in calculating physical observables. As $D$ increases, the nodal surface of PEPS $\left|\Psi_{\rm{P}}\right\rangle$ gets closer to the nodal surface of the true ground state. However, the time required for tensor contraction grows dramatically in $D^{12}$, and this process cannot be effectively accelerated.
On the other hand, the GFMC method is highly efficient and can be accelerated using parallel architectures, trading computational resources for time.
Once enough samples are collected, any observable can be calculated with almost no computational resources and time. This is in contrast to PEPS, which requires a high-dimensional tensor contraction for each calculation of an observable.

\section{Benchmark\label{sec:benchmark}}

We study the frustrated $J_1$-$J_2$ Heisenberg model to demonstrate the accuracy and efficiency of this hybrid method. The Hamiltonian of the model reads as
\begin{equation}
H = {J_1}\sum\limits_{\left\langle{i,j}\right\rangle}{{\mathbf S_i}\cdot{\mathbf S_j}} + {J_2}\sum\limits_{\left\langle\left\langle{i,j}\right\rangle\right\rangle}{{\mathbf S_i}\cdot{\mathbf S_j}},
\end{equation}
where $\mathbf S_i$ is the spin-$1/2$ angular momentum operator defined at the $i$-th site on a square lattice,  $\left\langle{...}\right\rangle$ and $\left\langle\left\langle{...}\right\rangle\right\rangle$ indicate the nearest and next-nearest neighbour sites, respectively. In this study, we set ${J_1}=1$, ${J_2}\geq 0$, i.e. both $J_1$ and $J_2$ are antiferromagnetic. The system contains $N = L \times L$ sites with PBCs.

The $J_1$-$J_2$ Heisenberg model has a rich phase diagram. When ${J_1}$ dominates, the system is in a Neel phase with antiferromagnetic (AFM) long-range order~\cite{Buonaura1998PRB, Sandvik1997PRB}, while the ground state manifests a well-established collinear AFM phase when ${J_2}$ dominates~\cite{neel1936proprietes,neel1932proprietes}.
Nevertheless, the intermediate regime $J_2/J_1\approx 0.5$ is a phase transition region with strong frustration and quantum fluctuations, and the physics remains a subject of considerable debate~\cite{singh1999dimer,takano2003nonlinear,zhang2003valence,mambrini2006plaquette,darradi2008ground,murg2009exploring,mezzacapo2012ground,jiang2012spin,Hu2013PRB,
morita2015quantum,wang2018critical,haghshenas2018u,ferrari2020gapless,hasik2021investigation,liu2022gapless,qian2024absence,Wang2013PRL,Liu2018PRB,
zhitomirsky1996valence,PhysRevLett.84.3173,PhysRevB.79.024409,PhysRevB.85.094407,PhysRevB.89.104415,PhysRevB.44.12050,Nomura2021PRX,Gong2014PRL,
Schulz1994Magnetic,Choo2019PRB,Zhang-PRB2024}.
Numerous studies have investigated the phases in this regime and drawn various conclusions, including the plaquette valence bond solid (VBS)~\cite{zhitomirsky1996valence,PhysRevLett.84.3173,PhysRevB.79.024409,PhysRevB.85.094407,PhysRevB.89.104415,takano2003nonlinear}, the columnar VBS~\cite{singh1999dimer,PhysRevB.44.12050,haghshenas2018u,Lin2024}, a gapless quantum spin liquid~\cite{zhang2003valence,Liu2018PRB,Wang2013PRL,Hu2013PRB}, and other proposals~\cite{mezzacapo2012ground,jiang2012spin,Gong2014PRL,morita2015quantum,wang2018critical,ferrari2020gapless,Nomura2021PRX,liu2022gapless,Zhang-PRB2024}.

\begin{figure}[htbp]
	\centering
	\includegraphics[scale=0.32]{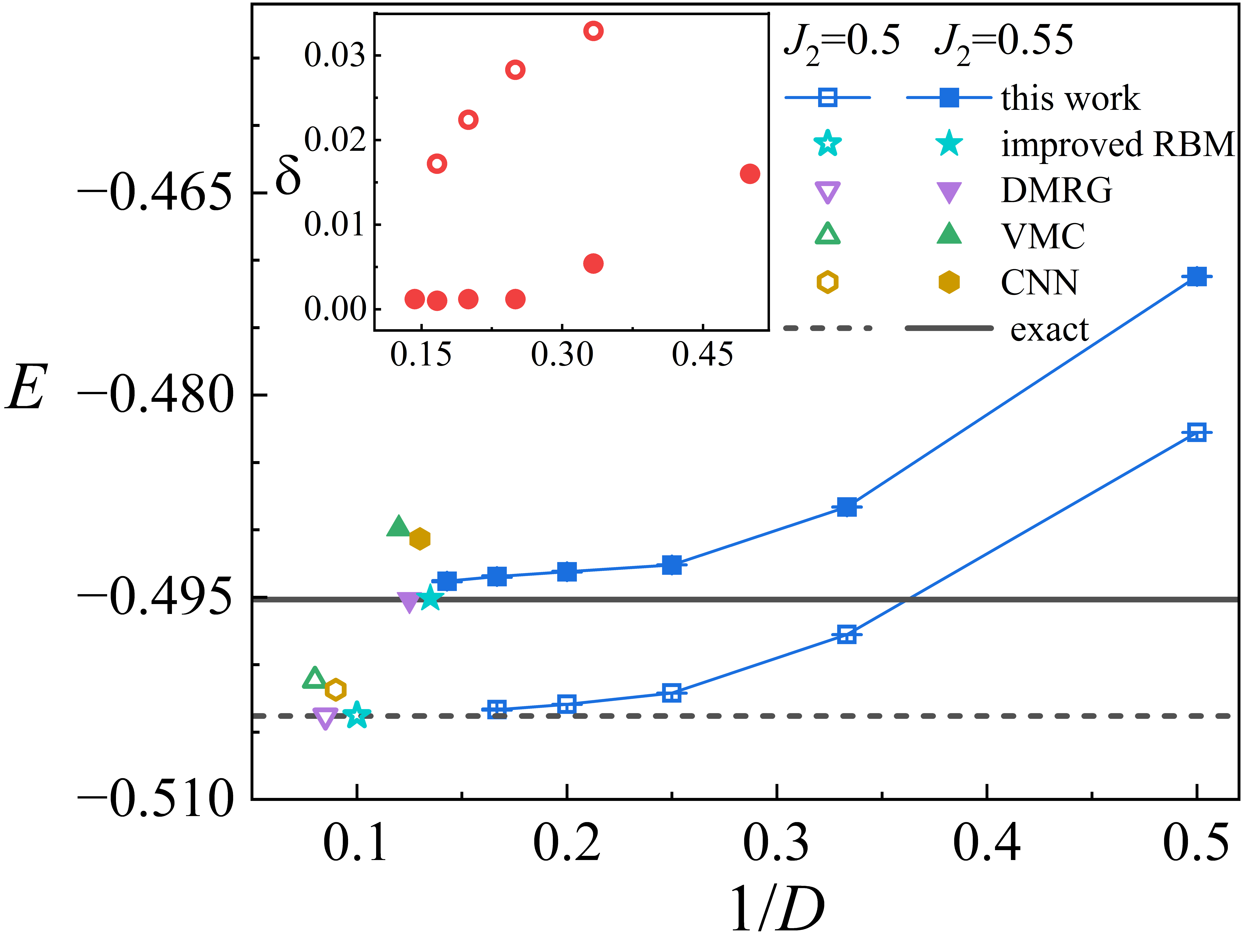}
	\caption{The ground-state energy $E$ as a function of $1/D$ on a $6\times 6$ square lattice with PBCs when $J_2 = 0.5$ (blue hollow squares) and $J_2 = 0.55$ (blue solid squares). The solid and dotted black horizontal lines represent the results of the exact diagonalization~\cite{Schulz1994Magnetic}. The extrapolated ones~\cite{Large-D} in the large-$D$ limit are close to the exact value to high precision. Results of the VMC~\cite{Hu2013PRB}, DMRG~\cite{Gong2014PRL}, CNN~\cite{Choo2019PRB}, improved RBM~\cite{Nomura2021PRX}, and the hybrid approach are shown as solid and hollow markers for $J_2 = 0.55$ and $J_2 = 0.5$, respectively. The inset shows the relative improvement $\delta =|E-E_{\rm{P}}|/|E_{\rm{P}}|$ as a function of $1/D$.}
	\label{fig:E_L6}
\end{figure}

We calculate the ground-state energies on $6\times6$ and $10\times10$ square lattices with $J_2 = 0.5$ and $J_2=0.55$, respectively.
In Fig.~\ref{fig:E_L6}, we report the ground-state energies obtained with trial states with different bond dimensions $D$ for a $6\times6$ lattice and benchmark our results with existing data in the literature.
For all data shown in Fig.~\ref{fig:E_L6}, solid markers represent results with $J_2 = 0.55$, and hollow markers represent the $J_2 = 0.5$ case.
We plot the energy $E$ as a function of $1/D$ with blue squares. It clearly shows that the energy can be systematically improved as $D$ becomes larger, and the extrapolated values~\cite{Large-D} in the large-$D$ limit are close to the exact ones~\cite{Schulz1994Magnetic} with high precisions, represented by the solid and dotted black horizontal lines.
We also compare the VMC~\cite{Hu2013PRB}, DMRG~\cite{Gong2014PRL}, CNN~\cite{Choo2019PRB}, and improved RBM~\cite{Nomura2021PRX} with the hybrid approach in Fig.~\ref{fig:E_L6}.
The relative improvement of the energy $E$ after the evolution with GFMC over the energy $E_{\rm{P}}$ of the PEPS trial wave function is defined as $\delta =|E-E_{\rm{P}}|/|E_{\rm{P}}|$, as shown in the inset of Fig.~\ref{fig:E_L6}.
For both $J_2$ values, the final energies $E$ obtained from GFMC are indeed lower than the initial values $E_{\rm{P}}$ provided by the PEPS wave function $|\Psi_{\rm{P}}\rangle$ as expected. 

\begin{table*}[htbp]
    \centering
	\caption{Comparison between the VMC, DMRG, improved RBM, Deep CNN, and the hybrid approach used in this paper on the $10\times10$ square lattice with PBCs.
    }
	\renewcommand\arraystretch{1}
	\setlength{\tabcolsep}{4mm}
	\begin{tabular}{|c|c|c|c|}
		\hline
		Method & $J_2=0.5$ & $J_2=0.55$ & Ref.\\
		\hline
		VMC & -0.49521(1) & -0.48335(1) & \cite{Hu2013PRB}\\
		\hline
		DMRG & -0.495530 & -0.485434 & \cite{Gong2014PRL}\\
		\hline
		this work (D=4) & -0.4954(6) & -0.4852(3) & -- \\
		\hline
		this work (D=5) & -0.4957(2) & -0.4853(4) & -- \\
		\hline
		improved RBM & -0.497629(1) & --  & \cite{Nomura2021PRX}\\
		\hline
        Deep CNN & -0.4976921(4) & --  & \cite{chen2023efficientoptimizationdeepneural}\\
		\hline
	\end{tabular}
	\label{tb:tb1}
\end{table*}

Table~\ref{tb:tb1} benchmarks our results with those from the state-of-the-art methods for the $10\times10$ lattice. As the results suggest, the ground-state energies obtained from the hybrid approach when $D = 5$ are comparable to those from DMRG~\cite{Gong2014PRL} and higher than those from improved RBM~\cite{Nomura2021PRX} and Deep CNN~\cite{chen2023efficientoptimizationdeepneural}.
These two methods are based on the neural-network quantum state optimization, and their performance can be found in Refs.~\cite{Choo2019PRB,Ledinauskas_2023,PhysRevResearch.2.033075,PhysRevB.98.104426,PhysRevB.103.035138,Wang_2024,PhysRevB.100.125131,PhysRevB.107.195115,
xiang2023density,9693260,PhysRevB.103.035138,Liang_2023,rende2023simple}.

\begin{figure}[htbp]
	\centering
	\includegraphics[scale=0.32]{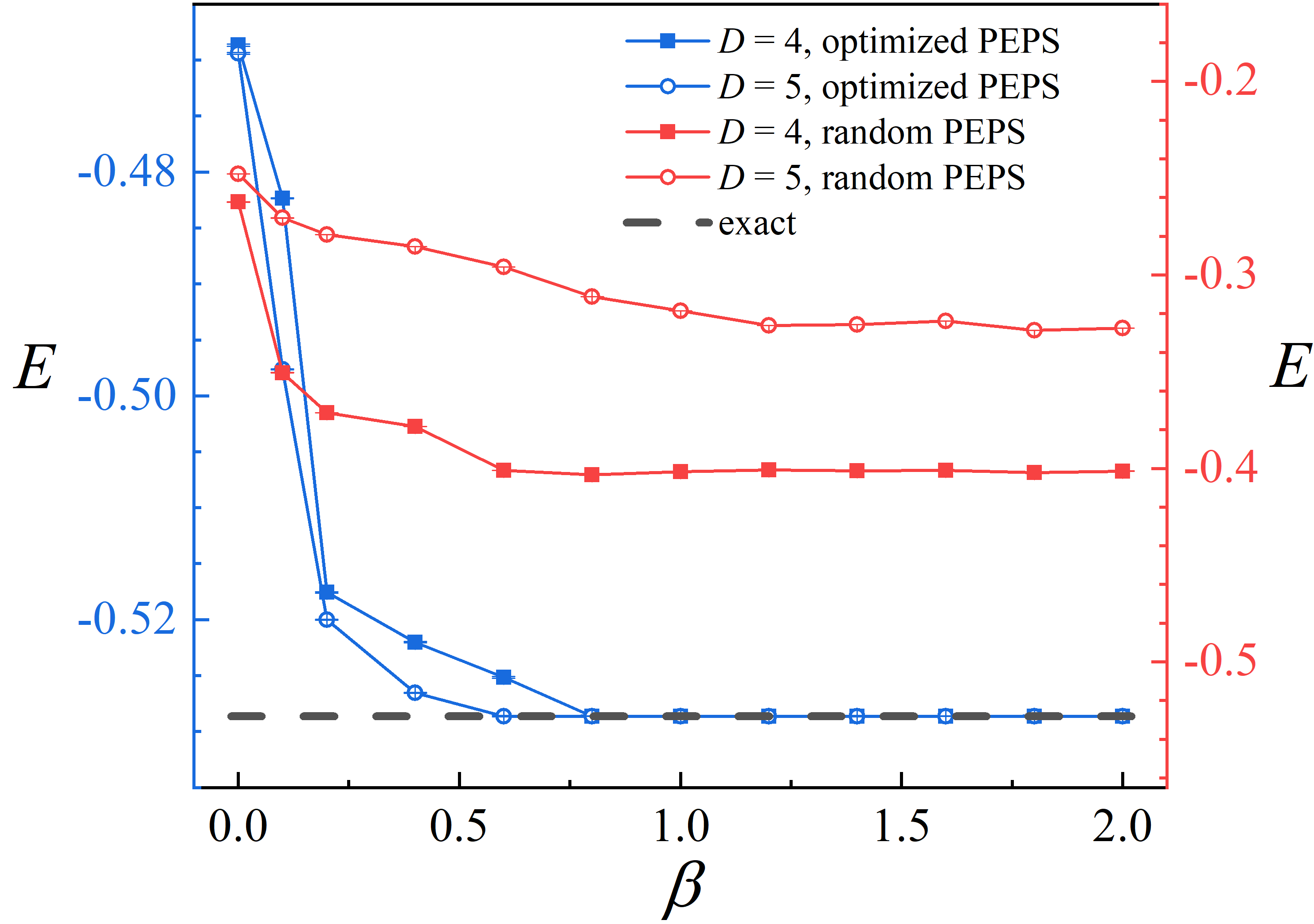}
	\caption{\add{The energies convergence with respect to imaginary time for different bond dimensions $D$ on a $4\times4$ square lattice for $J_2=0.5$. The black dashed line represents the exact ground-state energy, while the blue and red lines correspond to the results for optimized and random PEPS trial wave functions, respectively. Solid squares and hollow circles denote the results for $D=4$ and $D=5$.}}
	\label{fig:evolution}
\end{figure}

We show the convergence of the energy concerning the imaginary time for different $D$, as depicted in Fig.~\ref{fig:evolution} which report the results of adopting both optimized and random PEPS as trial wave functions in the GFMC method for $D=4$ and $D=5$ on a $4\times4$ square lattice for $J_2=0.5$.
The black dashed line represents the exact ground-state energy, while the blue and red lines correspond to the results for optimized and random PEPS trial wave functions. Solid squares and hollow circles denote the results for $D=4$ and $D=5$, respectively.
The figure shows that, for the optimized PEPS trial wave functions, the energy decreases from initial values of $-0.4686(7)$ for $D=4$ and $-0.4694(1)$ for $D=5$ to the exact value of $-0.528620$ as imaginary time evolves, with an error on the order of $10^{-5}$. The energies converge fast to the exact ground-state energy for the optimized PEPSs within $\beta \textless 1$. The convergence speed is slightly faster for $D=5$ compared to $D=4$. These indicate that both the optimized PEPS trial wave functions accurately capture the nodal surface of the ground state and a larger $D$ helps the convergence. In contrast, for the random PEPS trial wave functions, the energies fail to converge to the exact ground-state energy for both $D=4$ and $D=5$. This shows that the nodal surfaces of the random PEPS trial wave functions deviate significantly from that of the ground state.

Several prior methods have integrated TNS and MC techniques. For instance, Ref.\cite{Sandvik2007} employed MPS, while Ref.\cite{LiuWY2021} utilized PEPS in VMC simulations. Similar to our approach, the latter simplifies the PEPS double-layer network to a single-layer network, significantly enhancing contraction efficiency and leveraging the parallelism of Monte Carlo method. Both methods use Monte Carlo for acceleration, but with different approaches. The VMC method, based on the variational principle, avoids path integrals by relying on a positive probability distribution to circumvent the sign problem. In contrast, the GFMC method applies the fixed-node approximation to optimize the amplitude of the trial wave function, preserving more of the sign structure and potentially outperforming the use of PEPS alone.

\section{Summary\label{sec:summary}}
This paper presents a hybrid method that combines PEPS and GFMC. The PEPS, capturing the intrinsic entanglement properties of quantum states, provides a well-behaved trial state and a guiding wave function for GFMC. The subsequent imaginary-time evolution with GFMC is accelerated by parallelizing the importance sampling process, where the coefficient of each spin configuration is estimated by using HOTRG along one direction to contract a two-dimensional tensor network with bond dimension $D$. To demonstrate its effectiveness, we use this method to compute the ground-state energy of the $J_1$-$J_2$ Heisenberg model on a square lattice and benchmark it against other numerical methods. This hybrid approach achieves competitive accuracy in determining the ground-state energy, indicating its ability to accurately simulate the states in certain quantum frustrated spin systems.

Based on the results in this paper, we expect some potential improvements. They can be divided into two main types: finding a better initial PEPS for GFMC and exploring a better TNS ansatz than the PEPS used in GFMC initialization. One way to get better an initial PEPS is to apply the nested tensor network techniques~\cite{NTN2017,iTEBD2023} to increase the PEPS's available bond dimension, which is expected to serve as a better initial trial function, as suggested by the energy scaling behavior with the bond dimension. Another way is to directly optimize PEPS on a finite lattice instead of on a lattice in the thermodynamic limit, which could provide more accurate initial states for GFMC. Additionally, employing projected entangled simplex states~\cite{PESS2014,xiang2023density}, which have already been demonstrated with excellent performance in the frustrated spin systems, might give a better trial state for GFMC. We leave these implementations for future studies.

\section*{Program availability}

The code of this article has been published on GitHub at \href{https://github.com/lynn-cacher/gfmc.peps.code}{https://github.com/lynn-cacher/gfmc.peps.code} and is also openly available in Science Data Bank at https://doi.org/10.57760/sciencedb.j00113.00125. The README file provides a detailed description of the support required to run the program and the location of the configuration files for parameter settings.

\section*{Acknowledgment}
We are grateful to the discussion with Z.Y.Xie. Computational resources used in this study were provided by the Physical Laboratory of High-Performance Computing in Renmin University of China. This work was supported by the National Natural Science Foundation of China (Grants No. 11934020) and the Innovation Program for Quantum Science and Technology (Grants No. 2021ZD0302402). We thank the open-source library peps-torch~\cite{hasik2020peps} for sharing the code optimizing iPEPS with AD.

\end{document}